\documentclass[12pt]{article}
\usepackage{graphicx}
\usepackage[utf8]{inputenc}
\usepackage[english
]{babel}
\def\baselinestretch{1.5}
\begin{document}
\begin{center}
		\bf{Coherent states of parametric oscillators in the probability representation of quantum mechanics}\\
	\end{center}
	\bigskip
\footnotetext[2]{We dedicate this paper to the memory of
	Roy Jay Glauber, the great scientist and Nobel Prize Winner, on his
	first death anniversary, December 26, 2019. Ad~Memoriam of Roy
	Glauber and George Sudarshan is published in~\cite{CEWQO19,ICSSUR19}
	and is also available on
	\underline{link.springer.com/article/10.1007/s10946-019-09805-4} and
	\underline{www.mdpi.com/2624-960X/1/2/13}.}
	
	\begin{center} {\bf Vladimir N. Chernega$^1$, Olga V. Man'ko$^{1,2}$}
	\end{center}
	
	\medskip
	
	\begin{center}
		$^1$ - {\it Lebedev Physical Institute, Russian Academy of Sciences\\
			Leninskii Prospect 53, Moscow 119991, Russia}\\
		$^2$ - {\it Bauman Moscow State Technical University\\
			The 2nd Baumanskaya Str. 5, Moscow 105005, Russia}\\	
		Corresponding author e-mail: mankoov@lebedev.ru
	\end{center}

	\section*{Abstract}
Glauber coherent states of quantum systems are reviewed. We
construct the tomographic probability distributions of the
oscillator states. The possibility to describe quantum states by
tomographic probability distributions (tomograms) is presented on an
example of coherent states of parametric oscillator. The integrals
of motion linear in the position and momentum are used to explicitly
obtain the tomogram evolution expressed in terms of trajectories of
classical parametric oscillator	

	\medskip	
	
\noindent{\bf Keywords:} dichotomic random variables, probability
representation of quantum states, linear invariants, parametric oscillator,
coherent states, Wigner function.

\section{Introduction}
 In quantum mechanics, the states of a particle, e.g., of the
harmonic oscillator, are identified with the wave functions
$\psi(x,t)$ satisfying the Schr\"odinger evolution
equation~\cite{Sch26}, where $x$ is the oscillator position and $t$
is time. The energy levels and stationary states of the oscillator
and other systems are obtained by solving the stationary
Schr\"odinger equation $\hat H\psi_E(x)=E\psi_E(x)$, where $\hat H$
is the quantum system Hamiltonian. Among all solutions of the
evolution Schr\"odinger equations, there are specific
Gaussian-packet solutions for which the probability distribution
$P(x,t)=|\psi(x,t)|^2$ of the oscillator position at a given time
moment is described by the normal probability distribution of the
position $x$, with a given mean value $\bar x(t)$ and the dispersion
$\sigma(t)=\bar{x^2}(t)-(\bar x(t))^2$. Such packets were studied by
Schr\"odinger~\cite{Sch26}, and these oscillator states are similar
to classical oscillator states with fluctuating position and
momentum. In 1963, while studying the coherence properties of
photons, Roy~Glauber~\cite{Glauber1} introduced the notion and
terminology of {\it the field coherent states}; see also~[3--9].

For a single mode, the field is modeled by the quantum harmonic
oscillator, and the wave function $\psi_\alpha(x,t)$ of the harmonic
oscillator is the Gaussian packet satisfying the Schr\"odinger
evolution equation. Generic Gaussian states and entropic
inequalities for these states for multimode photon states were
studied in~\cite{MA-PS}.

The function $\psi_\alpha(x,t=0)$ is the eigenfunction of the photon
annihilation operator $\hat a\psi_\alpha(x)=\alpha\psi_\alpha(x)$,
where $\alpha$ is the complex number $\alpha=\alpha_1+i\alpha_2$;
see also~\cite{Scully,Schleih}.

The photon annihilation $\hat a$ and creation $\hat a^\dagger$
operators satisfy the bosonic commutation relations $[\hat a,\hat
a^\dagger]=1$. In the coherent states of the harmonic oscillator,
the Heisenberg uncertainty relations~\cite{Heis27} for dimensionless
position and momentum have the property $(\bar{x^2}-\bar
x^2)(\bar{p^2}-\bar p^2)=1/4$ and $\bar{x^2}-\bar x^2=\bar{p^2}-\bar
p^2=1/2$. The coherent state properties of the oscillator were
studied in~\cite{Glauber1,Glauber2,Glauber3}. The properties of the
coherent states of photons were also considered by
Sudarshan~\cite{Sud63}.

The aim of our work is to discuss the coherent states of the
parametric oscillator, i.e., of the oscillator with time-dependent
frequency $\omega(t)$. The Schr\"odinger evolution equation for such
oscillator was solved in~\cite{Husimi1953}. There is no energy
levels of the parametric oscillator, and the energy is not the
integral of motion. For a classical parametric oscillator, the
integral of motion, being quadratic in the position and momentum,
was found by Ermakov~\cite{Ermakov1880}. The quantum operator
quadratic in the position and momentum, being the integral of
motion, contains an explicit dependence on time in the Schr\"odinger
representation, as was found in~\cite{LewisResenfeld1969}. This
quantum integral of motion is an analog of the classical Ermakov
invariant, and it was used to find different solutions to the
Sch\"odinger equation in~\cite{LewisResenfeld1969}.

It was shown in~\cite{MalkinMankoTrifonov} that the parametric
oscillator has the linear (in the position and momentum) integrals of
motion $\hat A(t)$ and $\hat A^\dagger(t)$, which have the
commutation properties of bosonic annihilation and creation
operators, i.e., $[\hat A(t),\hat A^\dagger(t)]=1$. In view of what
we said above, one can extend the construction of Glauber coherent
states to the case of the parametric oscillator; see,
e.g.,~\cite{DodMalManko}). In view of developing the technique of
homodyne tomography of photon states~\cite{RaymerPRL93} based on the
relation between the Radon transform~\cite{Radon} of the Wigner
function~\cite{Wigner32} of the quantum system state with optical
tomogram, which is a fair probability distribution of the photon
quadrature found in~\cite{BerBer,VogRis}, the suggestion to identify
the quantum state with the probability distribution as a primary
object was done in~\cite{ManciniTombesiPLA}; see also the
review~\cite{MarmoPhysScr2010}.

The kinetic equation for the tomographic probability distribution,
which is the optical tomogram of the quantum state, with the wave function
obeying the Schr\"odinger evolution equation, was obtained
in~\cite{KorennoyJRLR,AmosovKorennoyMankoPhysRev}. This equation is
compatible with the kinetic equation for the symplectic tomogram of
quantum states introduced and studied
in~\cite{ManciniTombesiPLA,ManciniTombesiFoundPhys}. Such tomogram
exists and obeys the kinetic equation for the fair probability
distributions also in the case of a spin-1/2 particle, with the wave
function satisfying the Pauli equation~\cite{Pualeq}. Thus, in
addition to the review of Glauber's coherent states for the wave
function of the parametric oscillator, we consider the oscillator
coherent states in the probability representation of quantum
mechanics.

We present the evolution for the tomographic probability
distributions determining the oscillator states and construct the
probability distributions of the oscillator position in the form of
normal distribution with time-dependent parameters. The tomographic
probability distributions identified with the coherent states
satisfy the kinetic equations equivalent to the Schr\"odinger
equation for the wave function and the von-Neumann equation for the
density matrix of the parametric oscillator. As an application of
the formalism, we discuss the stimulated Raman scattering process in
the probability representation of quantum mechanics
in~\cite{JoptB2003,TchProcSPIE2006,TchJRLR2001}. The problem of
parametric oscillator was studied using different methods
in~[36--43].

This paper is organized as follows.

In Sec.~2, we present the method of linear integrals of motion to
find coherent states of a parametric oscillator. In Sec.~3, we give
a review of the conditional probability representation of quantum
states of the parametric oscillator. In Sec.~4, we construct the
joint probability distribution of three random variables for the
parametric oscillator in coherent states. In Sec.~5, we consider the
evolution of the parametric oscillator in the probability
representation  of quantum mechanics. Our conclusions and
prospectives are given in Sec.~6.

\section{Integrals of Motion of Parametric Oscillator
and Coherent States}   
The parametric oscillator has the
Hamiltonian
\begin{equation}\label{paros0}
\hat H=\frac{\hat p^2}{2}+\frac{\omega^2(t)\hat q^2}{2}.
\end{equation}
We assume the Planck constant $\hbar=1$, the oscillator mass $m=1$,
and frequency at time $t=0$, i.e., $\omega(0)=1$. The Schr\"odinger
equation for the oscillator wave function $\psi(x,t)$ in the
position representation
\begin{equation}\label{paros1}
i\frac{\partial \psi(x,t)}{\partial t}=-\frac{1}{2}
  \frac{\partial^2\psi(x,t)}{\partial t^2}+\frac{\omega^2(t)x^2}{2}\psi(x,t)
\end{equation}
was solved in~\cite{Husimi1953}, and various methods to study this
equation and its solutions were suggested
in~\cite{MalkinMankoTrifonov}. The method based on finding the
system's integrals of motion, which are operators quadratic in the
position and momentum, was used in~\cite{LewisResenfeld1969}.

The Ermakov integral of motion for a classical parametric oscillator
was found in~\cite{Ermakov1880}. The quantum version of the
classical Ermakov invariant depends on the solution of the classical
nonlinear
equation~\cite{Pinni,Schuch,SchuchCastanos,Rekamier,OccurOrtiz}.
Invariants, which are linear in the position and momentum operators,
were found in~\cite{MalkinMankoTrifonov}.

The time-dependent operators $\hat A(t)$ and $\hat A^\dag(t)$ of the
form
\begin{equation}\label{paros2}
\hat A(t)=\frac{i}{\sqrt2}\left(\epsilon(t)\hat p-\dot\epsilon(t)\hat q\right),
  \qquad \hat A^\dagger(t)=-\frac{i}{\sqrt2}\left(\epsilon^\ast(t)\hat p
  -\dot\epsilon^\ast(t)\hat q\right)
\end{equation}
are the linear integrals of motion satisfying the conditions
$$
\langle\psi(t)|\hat A(t)|\psi(t)\rangle=\langle \psi(0)|\hat A(0)|\psi(0)\rangle,
  \qquad \langle \psi(t)|\hat A^\dagger(t)|\psi(t)\rangle=\langle \psi(0)|\hat
  A^\dagger(0)|\psi(0)\rangle
$$
for function $\epsilon(t)$ satisfying the equation of motion for the classical
parametric oscillator $\ddot\epsilon(t)+\omega^2(t)\epsilon(t)=0$.

For initial conditions of the function $\epsilon(t)$ of the form
$\epsilon(0)=1$, $\dot\epsilon(t)=i$, the integrals of motion
(\ref{paros1}) and (\ref{paros2}) satisfy the commutation relation
\begin{equation}\label{paros3}
\left[\hat A(t),\hat A^\dagger(t)\right]=1,
\end{equation}
and these operators coincide for $t=0$ with the annihilation $\hat
a$ and creation $\hat a^\dagger$ operators of the harmonic
oscillator, i.e.,
$$
\hat A(0)=\hat a=\frac{1}{\sqrt2}\left(\hat q+i\hat p\right), \qquad
  \hat A^\dagger(0)=\hat a^\dagger=\frac{1}{\sqrt2}\left(\hat q-i\hat p\right).
$$

The coherent states of the parametric oscillator $|\alpha,t\rangle$
satisfying the equation $\hat
A(t)|\alpha,t\rangle=\alpha|\alpha,t\rangle$, where eigenvalues of
the integral of motion $\hat A(t)$ do not depend on time, are
obtained from an analog of the ground state of harmonic oscillator
$|0,t\rangle$ satisfying the Schr\"odinger equation and the condition
$\hat A(t)|0,t\rangle=0$ by means of the Weyl operator, which is the
integral of motion
\begin{equation}\label{paros4}
\hat D(\alpha)=\exp\left[\alpha\hat A^\dagger(t)-\alpha^\ast\hat A(t)\right].
\end{equation}
The coherent state reads
\begin{equation}\label{paros5}
|\alpha,t\rangle=\exp\left[\alpha\hat A^\dagger(t)-\alpha^\ast\hat A(t)\right]
|0,t\rangle.
\end{equation}
One can check that the function
\begin{equation}\label{paros6}
\psi_0(x,t)=\langle x|0,t\rangle=\frac{\pi^{-1/4}}{\sqrt{\epsilon(t)}}
  \exp\left(\frac{i\dot\epsilon(t)x^2}{2\epsilon(t)}\right)
\end{equation}
is the normalized solution to the Schr\"odinger
equation~(\ref{paros1}); for $t=0$, it is equal to the wave function
$\psi_0(x)=\pi^{-1/4}\exp(-x^2/2)$ of the oscillator ground state
satisfying the condition $\hat a\psi_0(x)=0$.

The Fock states of the parametric oscillator $|n,t\rangle$
satisfying the Schr\"odinger equation and the condition $\hat
A^\dagger(t)\hat A(t)|n,t\rangle=n|n,t\rangle$, where
$n=0,1,2,\ldots$, are given by the formula
\begin{equation}\label{paros7}
|n,t\rangle=\frac{\left(\hat A^\dagger\right)^n(t)}{\sqrt{n!}}|0,t\rangle.
\end{equation}
The coherent states of the parametric oscillator~(\ref{paros5}) are
expressed in terms of Fock states~(\ref{paros7}),
\begin{equation}\label{paros8}
|\alpha,t\rangle=e^{-|\alpha|^2/2}\sum_{n=0}^\infty\frac{\alpha^n}{\sqrt{n!}}
  |n,t\rangle.
\end{equation}
Since the coherent state $|\alpha,t\rangle$ is given by
Eq.~(\ref{paros5}), which provides the relation
$$|\alpha,t\rangle=e^{-|\alpha|^2/2}
\exp\left(\alpha\hat A^\dagger(t)\right)|0,t\rangle,$$ one has an explicit
expression for the wave function of the coherent state in the
position representation; it reads
\begin{equation}\label{paros9}
\psi_{\alpha}(x,t) = \langle x|\alpha,t\rangle=\psi_0(x,t)
  \exp\left(-\frac{|\alpha|^2}{2}+\frac{\sqrt2\alpha x}{\epsilon(t)}
  -\frac{\alpha^2\epsilon^\ast(t)}{2\epsilon(t)}\right).
\end{equation}
The wave function $\psi_n(x,t)=\langle x|n,t\rangle$ can be
obtained using the generating function for Hermite polynomials
\begin{equation}\label{paros10}
e^{-t^2+2t x}=\sum_{n=0}^\infty\frac{t^n}{n!}H_n(x),
\end{equation}
and formula (\ref{paros8}), where the parameter $\alpha$ is used to get
the coefficient in the series determining the vector $|n,t\rangle$
and consequently the wave function $\psi_n(x,t)$ in the
decomposition of the coherent-state wave function~(\ref{paros9}). We
obtain the wave function $\psi_n(x,t)$ in an explicit form as
follows:
\begin{equation}\label{paros11}
\psi_n(x,t)=\left(\frac{\epsilon^\ast(t)}{\epsilon(t)}\right)^{n/2}
  \frac{\psi_0(x,t)}{\sqrt{2^n n!}}H_n\left(\frac{x}{|\epsilon(t)|}\right).
\end{equation}
For $\omega(t)=1$ and $\epsilon(t)=e^{i t}$, the coherent-state wave
function becomes
\begin{equation}\label{paros12}
\psi_\alpha(x,t)=\frac{-(|\alpha|^2/{2})-{i t}/{2}}{\pi^{1/4}}
  \exp\left(-\frac{x^2}{2}+\sqrt2\alpha e^{-it}x-\frac{e^{-2i t}\alpha^2}{2}\right).
\end{equation}
The wave function (\ref{paros11}) has the standard form
\begin{equation}\label{paros13}
\psi_n(x,t)=e^{-i t(n+1/2)}\frac{e^{-x^2/2}}{\pi^{1/4}\sqrt{2^n n!}}H_n(x).
\end{equation}

\section{Tomographic Probability Representation of the Parametric Oscillator
States} 
The density matrix $\rho_\alpha(x,x',t)$ of coherent
states (\ref{paros5}) of the parametric oscillator has the Gaussian
form,
\begin{equation}\label{eq.p1'}
\rho_\alpha(x,x',t) = \psi_\alpha(x,t)\psi_\alpha^\ast(x',t) =
  \langle x |\hat\rho_\alpha(t)|x'\rangle.
\end{equation}
In~\cite{ManciniTombesiPLA}, the construction of the symplectic
tomographic probability representation of the system states with
continuous variables, like the oscillator, was proposed using the
invertible map of the state density operators at time $t=0$ onto
fair conditional probability distributions $w_\rho(X|\mu,\nu)$ of a
random variable (oscillator position) $-\infty \leq X\leq\infty$. It
depends also on the parameters $-\infty<\mu,\nu<\infty$
characterizing the reference frame in the phase space $(q,p)$, where
this position is measured. The map is given by the relation
\begin{equation}\label{eq.p2'}
w_\rho(X|\mu,\nu)=\mbox{Tr}\hat\rho\delta(X\hat1-\mu\hat q-\nu\hat p).
\end{equation}
The function is called the symplectic tomogram of the oscillator state.
The given formula can be used to express the density operator in
terms of the tomogram (probability distribution)
$w_\rho(X|\mu,\nu)$, i.e.,
\begin{equation}\label{eq.p3'}
\hat\rho=\frac{1}{2\pi}\int dX\, d\mu\,d\nu\, w_\rho(X|\mu,\nu)
  \exp i(X\hat1-\mu\hat q-\nu\hat p).
\end{equation}
In (\ref{eq.p2'}) and (\ref{eq.p3'}), operators $\hat q$ and $\hat
p$ are the position and momentum operators, respectively; also we
assume the Planck constant $\hbar=1$ as well as the oscillator mass
$m=1$.

For pure states $|\psi\rangle$, the expressions for the symplectic
tomogram can be given in terms of the fractional Fourier transform of
the wave function~\cite{MendesPLA},
\begin{equation}\label{eq.p4'}
w_\psi(X|\mu,\nu)=\frac{1}{2\pi|\nu|}\left|\int\psi(y)
  \exp\left(\frac{i\mu y^2}{2\nu}-\frac{i X y}{\nu}\right)d y\right|^2.
\end{equation}
Tomograms of pure and mixed states are nonnegative and satisfy the
normalization condition for arbitrary values of parameters $\mu$ and
$\nu$, i.e.,
\[\int w_\rho(X|\mu,\nu)\,d X=1.\]

The tomogram is related to the state's Wigner function
$W(q,p)$~\cite{Wigner32} given by the Fourier transform of the
density matrix $\rho(x,x')$,
\begin{equation}\label{eq.p5'}
W(q,p)=\int \rho\left(q+\frac{u}{2},q-\frac{u}{2}\right)e^{-i p u}d u.
\end{equation}
The relation is given by the Radon transform~\cite{Radon}
\begin{equation}\label{eq.p6'}
w(X|\mu,\nu)=\int W(q,p)\delta(X-\mu q-\nu p)\frac{d q\, d p}{2\pi}.
\end{equation}
The Wigner function can be reconstructed if the tomogram is known,
\begin{equation}\label{eq.p7'}
W(q,p) = \frac{1}{2\pi}\int w_\rho(X,\mu,\nu)\exp\left(i(X-\mu q-\nu p)\right)dX\,d\mu\,d\nu.
\end{equation}

For experimental study of photon states, optical tomograms $w^{\rm
(opt)}(X|\theta)$ measured by homodyne detectors, where $X$ is the
photon quadrature and $\theta$ is a local oscillator phase, are used
to reconstruct the Wigner function~\cite{RaymerPRL93}.

The symplectic tomogram determines the optical tomogram
$w_\rho(X|\theta)$ given by the relation
\begin{equation}\label{eq.p8'}
w_\rho^{\rm (opt)}(X|\theta) = \mbox{Tr}\left(\hat\rho\,\delta(X\hat1-\hat
q\cos\theta
  -\hat p\sin\theta)\right),
\end{equation}
which can be rewritten in terms of the Wigner function using the
Radon transform
\begin{equation}\label{eq.p9'}
w_\rho^{\rm (opt)}(X|\theta) = \int
W_\rho(q,p)\delta(X-q\cos\theta-p\sin\theta)
  \frac{dq\, dp}{2\pi}.
\end{equation}
This means that
\begin{equation}\label{eq.p10'}
w_\rho^{\rm (opt)}(X|\theta) =
w_\rho(X|\mu=\cos\theta,\nu=\sin\theta)
\end{equation}
and, in view of the Dirac delta-function property $\delta(\lambda
x)=|\lambda|^{-1}\delta(x)$, the optical tomogram determines the
symplectic tomographc probability distribution
\begin{equation}\label{eq.p11'}
w_\rho(X|\mu,\nu) = \frac{1}{\sqrt{\mu^2+\nu^2}}w_\rho^{\rm (opt)}
  \left(\frac{X}{\sqrt{\mu^2+\nu^2}}|\theta=\arctan\frac{\nu}{\mu}\right).
\end{equation}
For the parametric oscillator state with the wave function
(\ref{paros6}), the symplectic tomographic probability distribution
is the normal distribution of a random variable $X$; it has the form
\begin{equation}\label{eq.p12'}
w_0(X|\mu,\nu,t) = \frac{1}{\sqrt{2\pi\sigma(\mu,\nu,t)}}
  \exp\left(-\frac{X^2}{2\sigma(\mu,\nu,t)}\right).
\end{equation}
Here, the dispersion parameter reads
\begin{equation}\label{eq.p13'}
\sigma(\mu,\nu,t) = \mu^2\frac{|\epsilon(t)|^2}{2}+\nu^2\frac{|\dot\epsilon(t)|^2}{2}
  +2\mu\nu\sigma_{q p}(t).
\end{equation}
This expression follows from the relation determined by the contribution
of the Dirac delta-function term in the density operator
(\ref{eq.p3'})
\begin{equation}\label{eq.p14'}
X\hat1=\mu\hat q+\nu\hat p,
\end{equation}
which provides the equality
\begin{equation}\label{eq.p15'}
\langle X^2\hat1\rangle=\mu^2\langle\hat q^2\rangle_0+\nu^2\langle\hat p^2\rangle_0
+2\mu\nu\left\langle\frac{\hat q\hat p+\hat p\hat q}{2}\right\rangle_0.
\end{equation}
For the parametric oscillator state with the wave function
(\ref{paros6}), one has
$$
\langle\hat q\rangle=0,\qquad\langle\hat p
\rangle=0,\qquad\langle\hat
q^2\rangle_0=\frac{|\epsilon(t)|^2}{2},\qquad \langle\hat
p^2\rangle_0=\frac{|\dot\epsilon(t)|^2}{2},$$ and the covariance
term satisfies the relation depending on the correlation coefficient
\begin{equation}\label{eq.p16'}
r^2=\frac{\sigma^2_{q p}}{\langle\hat q^2\rangle_0\langle\hat
p^2\rangle_0},\qquad\sigma_{q p}=\left\langle\frac{\hat q\hat p+\hat
p\hat q}{2}\right\rangle_0,\qquad
\langle \hat q^2\rangle_0\langle \hat p^2\rangle_0=\frac{1}{4}\cdot\frac{1}{1-r^2}\,.
\end{equation}
This equality means that the state~(\ref{paros6}) provides the bound
in the Schr\"odinger--Robertson~\cite{Sch30,Rob29} uncertainty
relation.

Thus, we have the following property of the quantum parametric
oscillator state~(\ref{paros6}). The variances and covariances of
these oscillator state are determined by the solution $\epsilon(t)$
and $\dot\epsilon(t)$ for the classical parametric oscillator
motion. The correlation coefficient $r(t)$, being dependent on time
$t$, is expressed also in terms of the trajectories $\epsilon(t)$
and $\dot\epsilon(t)$ of the classical parametric
oscillator~\cite{KurmisDodManPLA},
\begin{equation}\label{eq.p17'}
r^2(t)=1-|\epsilon(t)\dot\epsilon(t)|^{-2}.
\end{equation}

For the coherent state of the parametric oscillator with the wave
function (\ref{paros9}), the tomographic probability distribution
$w_\alpha(X|\mu,\nu,t)$ has the form of the normal probability
distribution
\begin{equation}\label{eq.p18'}
w_\alpha(X|\mu,\nu,t)=\frac{1}{\sqrt{2\pi\sigma_\alpha(\mu,\nu,t)}}
  \exp\left(-\frac{(X-\bar X_\alpha(\mu,\nu,t))^2}{2\sigma_\alpha(\mu,\nu,t)}\right),
\end{equation}
where
\begin{equation}\label{eq.p19'}
\bar X_\alpha(\mu,\nu,t)=\mu\langle\hat q\rangle_\alpha+\nu\langle\hat p\rangle_\alpha
\end{equation}
and
\begin{equation}\label{eq.p20'}
  \langle\hat q\rangle_\alpha=\sqrt2\Re(\alpha\epsilon^\ast(t)),\qquad
  \langle\hat p\rangle_\alpha=\sqrt2\Im(\alpha\dot\epsilon^\ast(t)).
\end{equation}
The parameter $\sigma_\alpha(\mu,\nu,t)=\sigma(\mu,\nu,t)$ is given
by (\ref{eq.p13'}).

Thus, the fair probability distribution~(\ref{eq.p18'}) describes
coherent states of the parametric oscillator, and this probability
distribution contains complete information on the state.

The optical tomographic probability distribution $w_\alpha^{\rm
(opt)}(X|\theta)$ of the coherent state, which can be measured by
the homodyne detector, has the form~(\ref{eq.p18'}) with the parameters
$\mu=\cos\theta$ and $\nu=\sin\theta$. This means that, for the
parametric oscillator, the optical tomogram of the coherent state
reads
\begin{equation}\label{eq.p21'}
w_\alpha(X|\theta) = \frac{1}{\sqrt{2\pi\sigma_\alpha(\theta,t)}}
  \exp\left[-\frac{(X-\bar X(\theta,t))^2}{2\sigma_\alpha(\theta,t)}\right],
\end{equation}
where the mean photon quadrature is
$$
\bar X(\theta,t) = \cos\theta\sqrt2\Re(\alpha\epsilon^\ast(t))
  +\sin\theta\sqrt2\Im(\alpha\dot\epsilon^\ast(t))
$$
and
$$
\sigma_\alpha(\theta,t) = \cos^2\theta\frac{|\epsilon(t)|^2}{2}
+\sin^2\theta\frac{|\dot\epsilon(t)|^2}{2}+\sin2\theta\left(|\epsilon(t)\dot\epsilon(t)|^2-1\right)^{1/2}.
$$

The physical properties of coherent states of the parametric
oscillator depend on frequency $\omega(t)$. Varying the frequency,
one can create both the squeezing phenomenon, i.e., $|\epsilon(t)|^2<1$
or $|\dot\epsilon(t)|^2<1$, as well as the correlation phenomenon,
when $|\epsilon(t)\dot\epsilon(t)|^2>1$. The uncertainty
relation~\cite{Sch30,Rob29} guarantees that, for the classical complex
trajectory $\epsilon(t)$, one has the inequality
$|\epsilon(t)\dot\epsilon(t)|^2\geq1.$ One can conjecture that, in
the case of correlated coherent states discussed
in~\cite{KurmisDodManPLA}, the squeezing phenomenon of Gaussian
states can also take place.

\section{Conditional and Joint Probability Distributions Determining the Oscillator's Coherent States}
The symplectic tomographic probability distribution
$w_\psi(X|\mu,\nu)$ of the parametric oscillator state with the wave
function $\psi(x)$ is determined in terms of the fractional Fourier
transform of the wave function (\ref{eq.p4'})~\cite{MendesPLA},
where $X$ is the oscillator position measured in the reference frame
of the oscillator phase space determined by real parameters $\mu$
and $\nu$; $-\infty<\mu,\nu<\infty$.

In the case of the classical parametric oscillator, one has the relation
$X=\mu q+\nu p$; for $\mu=s\cos\theta$ and $\nu=s^{-1}\sin\theta$,
the reference frame parameters $s$ and $\theta$ provide the scale
changes of the form $q\rightarrow q'=s q$ and $p\rightarrow
p'=s^{-1}p$, along with the rotation of the axes $q'\rightarrow
X=\cos\theta \,q'+\sin\theta \,p'$ and $p'\rightarrow{\cal
P}=\sin\theta \,q'+\cos\theta \,p'$. The tomogram does not depend on
the variable ${\cal P}$.

The dependence on the parameters $\mu$ and $\nu$ of the symplectic
tomogram $w_\psi(X|\mu,\nu)$ provides the interpretation of the
tomographic conditional probability distribution of the system
position in the given reference frame. Using the Bayes' formula, the
joint probability distribution $w_\psi(X,\mu,\nu)$ with the
corresponding conditional probability distribution (\ref{eq.p4'})
can be introduced~\cite{EntropyMMankoVManko}
\begin{equation}\label{3.2}
w_\psi(X,\mu,\nu)=w_\psi(X|\mu,\nu){\cal P}
(\mu,\nu),
\end{equation}
where ${\cal P}(\mu,\nu)$ is an arbitrary normalized marginal
probability distribution $0\leq{\cal P}(\mu,\nu)\leq1$, i.e.,
\begin{equation}\label{3.3}
\int{\cal P}(\mu,\nu)\,d\mu \,d\nu=1.
\end{equation}
For example, the function ${\cal P}(\mu,\nu)$ can be chosen as
the normal distribution  ${\cal P}(\mu,\nu)=\frac{1}{\pi}\exp(-\mu^2-\nu^2)$.

For a harmonic oscillator with frequency $\omega=1$ and $m=1$
described by the wave function of the coherent
state~(\ref{paros12}), the tomogram $w_\alpha(X,\mu,\nu,t=0)$ is
given as the normal joint probability distribution of three random
variables
\begin{equation}\label{17'}
w_\alpha(X,\mu,\nu,t=0)=\frac{1}{\sqrt{2\,\pi\sigma}}
  \exp\left[-\frac{(X-\bar X)^2}{2\,\sigma}\right]{\cal P}(\mu,\nu),
\end{equation}
where $\bar
X=\sqrt2\left(\mu\mbox{Re}\,\alpha+\nu\mbox{Im}\,\alpha\right)$ and
$\sigma=(\mu^2+\nu^2)/2.$

For coherent state, the parametric oscillator tomogram
$w_\alpha(X|\mu,\nu,t)$ determines the joint probability
distribution
\begin{equation}\label{17''}
w_\alpha(X,\mu,\nu,t) =
\frac{\exp(-\mu^2-\nu^2)}{\pi\sqrt{2\,\pi\sigma_p(t)}}
  \exp\left[-\frac{(X-\bar X(t))^2}{2\,\sigma_p(t)}\right],
\end{equation}
where $\bar X=\mu\bar q(t)+\nu\bar p(t)$,
$\bar{q}(t)=\sqrt2\,\mbox{Re}\,[\alpha\epsilon^\ast(t)]$, and
$\bar{p}(t)=\sqrt2\,\mbox{Re}\,[\alpha\dot\epsilon^\ast(t)]$.

The joint probability distribution (\ref{17''}) of three random
variables $X$, $\mu$, and $\nu$ provides the possibility to
reconstruct tomogram (\ref{eq.p2'}) using the Bayes' formula
\begin{equation}\label{17'''}
w_\rho(X|\mu,\nu)=w_\alpha(X,\mu,\nu){\cal P}^{-1}(\mu,\nu).
\end{equation}

\section{The Parametric State Evolution in the Probability Representation}  
For a given Hamiltonian of the parametric oscillator, the unitary
evolution of the state vector $|\psi(t)\rangle=\hat
U(t)|\psi(0)\rangle$ provides the evolution of the density operators
$\hat\rho_\psi(t)=\hat U(t)|\psi(0)\rangle\langle\psi(0)|\hat
U^\dagger(t)$ and the evolution of the tomographic probability
distribution of the form
\begin{equation}\label{eq.q1''}
w(X|\mu,\nu,t) = \int G(X,\mu,\nu,X',\mu',\nu',t)w(X'|,\mu',\nu',t=0)\,dX'\,d\mu'\,d\nu'.
\end{equation}
In this section, we demonstrate that the evolution is given, using a
specific change of the variables $X,\mu,\nu\rightarrow
X(t),\mu(t),\nu(t)$ determined by the classical trajectories
$\epsilon(t)$ and $\dot\epsilon(t)$.

In view of the properties of the Dirac delta-function $\delta(\lambda
y)=|\lambda|^{-1}\delta(y)$, the tomogram $w(X|\mu,\nu,t)$ for
arbitrary time $t$ has the property
\begin{equation}\label{eq.q2''}
w(\lambda X|\lambda\mu,\lambda\nu,t)=|\lambda|^{-1}w(X|\mu,\nu,t).
\end{equation}
The density operator $\hat\rho(t)$ of an arbitrary state of the
parametric oscillator evolves according to the following form of the
solution
\begin{equation}\label{eq.q4''}
\hat\rho=\hat U(t)\hat\rho(0)\hat U^\dagger(t)
\end{equation}
of the von Neumann equation
\begin{equation}\label{eq.q3''}
\dot{\hat\rho}(t)+i[\hat H(t),\hat\rho(t)]=0.
\end{equation}
Here, the unitary operator $\hat U(t)$ is the solution of the
Schr\"odinger equation
\begin{equation}\label{eq.q5''}
i\frac{\partial\hat U(t)}{\partial t}=\hat H(t)\hat u(t),\quad \hat U(0)=1.
\end{equation}
Calculating the tomographic probability distribution
$w_\rho(X|\mu,\nu,t)$, in view of (\ref{eq.p2'}), we arrive at
\begin{equation}\label{eq.q6''}
w_\rho(X|\mu,\nu,t)=\mbox{Tr}(\hat U(t)\hat\rho(0)\hat U^\dagger(t)\delta\left(X\hat 1-\mu\hat q-\nu\hat p)\right).
\end{equation}
Using the relation
\begin{equation}\label{eq.q6'''}
\hat U^\dagger(t)\delta(X\hat1-\mu\hat q-\nu\hat p)\hat U(t)=\delta(X\hat1-\mu\hat q_H(t)-\nu\hat p_H(t),
\end{equation}
where operators $\hat q_H(t)$ and $\hat p_H(t)$ are the position and
momentum operators of the parametric oscillator in the Heisenberg
representation, we obtain the tomogram $w_\rho(X|\mu,\nu,t)$ as
follows:
\begin{equation}\label{eq.q7''}
w_\rho(X|\mu,\nu,t)=w_\rho(X|\mu_H(t),\nu_H(t),t=0).
\end{equation}
Parameters $\mu_H(t)$ and $\nu_H(t)$ are linear combinations of the
parameters $\mu$ and $\nu$ with coefficients depending on the
functions $\epsilon(t)$ and $\dot\epsilon(t)$. The integrals of
motion $\hat{A}(t)$ and $\hat A^\dagger(t)$ (\ref{paros2}),
satisfying comutation relations (\ref{paros3}) and the conditions
\begin{equation}\label{eq.q9''}
\frac{d\hat A(t)}{d t}+i[\hat H(t),\hat A(t)]=0,\qquad
\frac{d\hat A^\dagger(t)}{d t}+i[\hat H(t),\hat A^\dagger(t)]=0
\end{equation}
provide the possibilities to obtain the Heisenberg position operator
$\hat q_H(t)$ and momentum operator $\hat p_H(t)$ satisfying the
equations
\begin{equation}\label{eq.q10''}
\frac{\partial\hat q_H(t)}{\partial t}-i[\hat H(t),\hat q_H(t)]=0,\qquad
  \frac{\partial\hat p_H(t)}{\partial t}-i[\hat H(t),\hat p_H(t)]=0,
\end{equation}
as the linear combination of operators $\hat q$ and $\hat p$. This
means that we obtain the following transform of the Dirac
delta-function:
\begin{equation}\label{eq.q11''}
\delta(X\hat1-\mu\hat q_H(t)-\nu\hat p_H(t))=\delta(X\hat1-\mu_H(t)\hat q-\nu_H(t)\hat p).
\end{equation}
Finally, we have explicit expressions for operators $\hat q_H(t)$
and $\hat p_H(t)$ in terms of complex functions $\epsilon(t)$ and
$\dot\epsilon(t)$; they read
\begin{eqnarray}
\hat q_H(t)&=&\frac{1}{2}\{\epsilon^\ast(t)(\hat q+i\hat p)+\epsilon(\hat q-i\hat
p)\},\nonumber\\[-2mm]
\label{eq.q12''}\\[-2mm]
\hat p_H(t)&=&\frac{1}{2}\{\dot\epsilon^\ast(t)(\hat q+i\hat p)+\dot\epsilon(\hat q-i\hat p)\}.
\nonumber
\end{eqnarray}
In view of these explicit expressions, we arrive at
\begin{eqnarray}
\mu_H(t)&=&\frac{1}{2}\{\mu[(\epsilon^\ast(t)+\epsilon(t)]+\nu[\dot\epsilon^\ast(t)+\epsilon(t)]\},
\nonumber\\[-2mm]
\label{eq.q13''}\\[-2mm]
\nu_H(t)&=&\frac{i}{2}\{\mu[(\epsilon^\ast(t)-\epsilon(t)]+\nu[\dot\epsilon^\ast(t)-\epsilon(t)]\}.
\nonumber
\end{eqnarray}
Thus, for an arbitrary state of the parametric oscillator
$\hat\rho(0)$, the initial tomogram $w_{\rho(0)}(X|\mu,\nu)$ becomes
the tomographic probability distribution with the time dependence
given by formula~(\ref{eq.q7''}), where the parameters $\mu_H(t)$ and
$\nu_H(t)$ are given by (\ref{eq.q13''}). Such kind of
tomographic probability evolution takes place for arbitrary systems
with Hamiltonians quadratic in the position and momentum.

\section{Conclusions}    
To conclude, we point out the main results of our work.

We reviewed the known solution to the Schr\"odinger equation for
parametric oscillator. We constructed the probability distributions,
which can be identified with coherent states of a parametric
oscillator. The dynamics of symplectic and optical tomographic
probability distributions for the states of a quantum parametric
oscillator is expressed in terms of classical trajectories of the
classical parametric oscillator. Coherent states of a quantum
parametric oscillator, which describe the phenomenon of squeezing
and correlation of the oscillator's position and momentum, are
considered in the probability representation of quantum mechanics,
and the optical and symplectic tomograms of the oscillator are
obtained explicitly. Different aspects of the tomographic approach
to studying photon states, oscillator states, and qubit states were
considered in~[49--51]. 

The tomographic probability distributions can also describe
classical oscillator states identified with the probability
densities in the phase space. The classical oscillator states with
Gaussian probability density in the phase space have symplectic and
optical tomograms, which are normal probability distributions
$w_{\rm cl}(X|\mu,\nu)$ as in the case of quantum parametric
oscillator considered in this work. But the set of such states for
a classical parametric oscillator contains tomograms violating the
Schr\"odinger--Robertson uncertainty relation. If one reconstructs
the formal density operator, using such tomographic probability
distribution of the classical parametric oscillator state with
Gaussian tomogram and Gaussian probability density in the phase
space, the formal  density operator will have negative eigenvalues.
The relation of tomograms of classical and quantum oscillators, as
well as the case of multimode parametric oscillator and its coherent
states, will be discussed in future publications.

\end{document}